\documentclass[12pt]{iopart}  
\usepackage{graphicx}
\usepackage{citesort}
\usepackage{amssymb}
\usepackage{textcomp,libertine}
\usepackage{color}
\usepackage[normalem]{ulem}

\begin{document}

\title[Correlation between magnetism and lattice dynamics for cubic FeGe under pressure]
      {Correlation between magnetism and lattice dynamics for cubic FeGe under pressure}

\author{R-A Tonacatl-Monez$^1$, R Heid$^2$, and O De la Pe\~na-Seaman$^{1}$}

\address{$^1$ Instituto de F\'isica ``Ing. Luis Rivera Terrazas'', Benem\'erita 
         Universidad Aut\'onoma de Puebla, Av. San Claudio \& Blvd. 18 Sur, 
				 Ciudad Universitaria, C.P. 72570, Puebla, Puebla, M\'exico}
\address{$^2$ Institut f\"{u}r QuantenMaterialien und Technologien, 
         Karlsruher Institut f\"{u}r Technologie (KIT), D-76021 Karlsruhe, Germany}

\eads{omar.seaman@correo.buap.mx}

\begin{abstract}
This first-principles study investigates the structural, electronic, lattice 
dynamical properties, and electron-phonon coupling in ferromagnetic cubic B20 
FeGe under applied pressure. The implemented spin-scaling exchange-correlation 
(ssxc) approach allowed to modify the magnetic moment and ferromagnetic phase 
energetics using a single scaling parameter, thereby yielding an adjustment of 
the critical pressure ($p_c$) to its experimental value. 
The ssxc scheme resulted in a subtle energy shift of the electronic bands in 
the spin-up channel, and reduced the magnetic moment, bringing it closer to the 
experimentally reported value.
Application of the ssxc approach to phonon dispersion and electron-phonon 
interaction resulted in a slight mitigation of the pronounced softening and 
large linewidths of the lowest-frequency acoustic branch close to the 
$R$-point, typically observed with standard DFT calculations. With increasing 
pressure, phonon anomaly and linewidths diminish significantly and practically 
disappear at $p_c$ and beyond. This trend parallels the pressure dependence of 
the magnetic moment. A comparative analysis of the electronic joint density of 
states with the phonon linewidths revealed that the momentum dependence of 
linewidths around the $R$-point closely follow the momentum dependence of the 
electron-phonon matrix elements.
This indicates that the correlation between magnetic moment and linewidths 
under applied pressure originates from the electron-phonon matrix elements, 
presenting a distinct scenario compared to other B20 family members, where 
nesting plays a more dominating role.
\end{abstract}

\vspace{2pc}

\submitto{\JPCM}
%

\section{Introduction}

Transition metals (Mn, Fe, Co) together with group 14 elements (Si, Ge) form 
compounds that can crystallize in two polymorphs with hexagonal B35 (space 
group P6/mmm, $\# 191$) and cubic B20 structure (space group P2$_1$3, $\# 
198$), that interestingly lacks inversion symmetry. 
These systems are of interest for technological development due to their 
magnetic properties, such as helical spin ordering and the presence of 
skyrmionic phases for the B20 structures, which show great potential for 
spintronics applications \cite{kise,kana1}. Notable examples of B20 systems 
that have attracted significant interest include FeSi \cite{eo}, MnSi 
\cite{ishi}, FeGe \cite{grigo1,wang}, and the Fe$_{1-x}$Co$_x$Si solid solution 
\cite{kana2}.
\\

In particular, FeSi has been under intense investigation due to the 
temperature-dependent behavior of its electronic, transport, and magnetic 
properties \cite{schle,buschi}. Furthermore, a strong temperature dependence of 
its phonon spectra has been observed, specifically an anomalous softening of 
low-frequency phonon modes along the $[111]$ high-symmetry direction (close to 
the R-point), around $100$~K, which corresponds to a crossover from a 
low-temperature non-magnetic ground state to a state with an induced 
paramagnetic moment \cite{jacca,mandru}. This phenomenon, along with drastic 
changes in the related phonon linewidths that follow the observed magnetization 
trends with temperature, clearly indicates a correlation between magnetism and 
lattice dynamic properties in FeSi \cite{menzel,delaire,krann}.
This interplay between magnetic order and lattice dynamics is not unique to 
FeSi. For instance, in MnSi, substitutional doping with Fe (Mn$_{1-x}$Fe$_x$Si) 
leads to the suppression of its helical magnetic order, reaching a magnetic 
critical point at a critical concentration of $x_c=0.17$ \cite{bannen}. 
Interestingly, this system also exhibits a significant softening and broadening 
of a low-frequency phonon mode along the $[111]$ direction with an increasing 
Fe-content, a phenomenon linked to changes in the Fermi surface geometry 
\cite{khan}. \\

A similar behavior of the low-frequency phonon modes along the same direction 
of FeSi and Mn$_{1-x}$Fe$_x$Si (both at its magnetic phase) is also observed in 
ferromagnetic first-principles calculations for cubic FeGe \cite{wilhm1}. 
The B20 cubic polymorph is stable below $853$~K, while above this temperature 
the hexagonal structure is preferred \cite{stolt}. In the B20 structure, FeGe 
has a long-range helical spin structure, along the [111] direction, below a 
critical temperature of $T_c \sim 280$~K \cite{grigo1}. 
Near $T_c$ its characteristic helical magnetic order transits to a conical one 
as the applied magnetic field raises, reaching a skyrmion phase (swirling spin 
texture) for a specific region of magnetic field and temperature \cite{yu}.
Unlike FeSi, FeGe experimentally exhibits a reduction of the magnetic moment 
(from $3.928 \pm 0.007 \mu_B$ per unit cell) as the temperature increases up to 
$T_c$, where it undergoes a magnetic to non$-$magnetic phase transition 
\cite{spen,wilhm2}. However, neither experimental nor theoretical work exists 
addressing the question, if the phonon softening predicted for the 
ferromagnetic phase also persists into the non-magnetic high-temperature phase.
In addition to modifying the FeGe magnetic phase by temperature, as in FeSi, or 
by doping, like Mn$_{1-x}$Fe$_x$Si, experimental studies show a reduction of 
the magnetic moment as a function of applied pressure. In particular, the 
observed long-range helical magnetism (typical of B20 materials) disappears at 
approximately $19$~GPa, reaching an inhomogeneous chiral spin state which 
finally changes to the paramagnetic state at a critical pressure ($p_c$) value 
of $28.5$~GPa \cite{barla}. There have been attempts to reproduce the observed 
$p_c$ value through first-principles studies, yielding a wide range of values, 
from $10$~GPa to $40$~GPa with local-density approximation and  
generalized gradient approximation (PW91) functionals, respectively 
\cite{neef}. Clearly, such reported values differ from the experimentally 
determined one, although the FeGe structural ground-state properties are 
reasonably well described. \\

Thus, the aim of this work is twofold for FeGe: First, analyze a different 
methodological approach to properly describe the critical pressure at which the 
magnetic transition takes place and sense its influence on the electronic and 
lattice dynamical properties, and second, to analyze the evolution of the 
lattice dynamics (phonon frequencies and linewidths) as a function of pressure. 
This analysis aims to determine if there is any correlation between magnetism 
and lattice dynamics, similar to what is observed in FeSi or 
Mn$_{1-x}$Fe$_x$Si, by using applied pressure, instead of temperature or doping 
content, as a method to modulate magnetism in the system.

\section{Methodology}
First-principles calculations were performed using density functional theory 
(DFT) \cite{hohen,kohn} implemented in the mixed-based pseudopotential method 
(MBPP) \cite{mbpp} and using the generalized gradient approximation by 
Perdew-Burke-Ernzerhof (GGA-PBE) \cite{perdw} for the exchange and correlation 
functional, neglecting spin-orbit coupling. The non-magnetic (NM) calculations 
were considered to model the paramagnetic state. We are aware that the 
NM state does not take into account the localized magnetic moments present in 
the paramagnetic one. However, it represents an approximate description of the 
non-polarized electronic system. The magnetic phase was approximated by a 
collinear spin-polarized , i.e. ferromagnetic (FM) phase, since the FeGe helical 
spin structure possesses a long pitch-period in comparison with the unit cell 
(approximately 150 times larger) \cite{lebe}.
Norm-conserving pseudopotentials for Fe and Ge were constructed using the 
Vanderbilt scheme \cite{vand}, including semicore states Fe $3s$ and $3p$ in 
the valence region. Furthermore, $s$, $p$, and $d$ 
local functions at Fe sites were added to the plane-wave expansion up to a 
kinetic energy of $24$~Ry. For the integration of the Brillouin zone, the 
Monkhorst-Pack special $k$-point set technique was used, with a Gaussian 
smearing of $0.1$~eV and a grid of $16 \times 16 \times 16$ for the crystal 
structural optimization, and a $48 \times 48 \times 48$ mesh for the electronic
properties. \\

Phonon properties, that is, phonon dispersion and phonon density of states, 
were calculated via density functional perturbation theory (DFPT) 
\cite{louie,baron}, implemented in the MBPP code \cite{heid2}. Complete phonon 
dispersions were obtained via standard Fourier interpolation of dynamical 
matrices calculated on a $4 \times4 \times 4$ $q$-point mesh, on basis of a
$16 \times 16 \times 16$ $k$-point grid. The differences in interpolation 
(phonon dispersion) were minimal (less than $0.4$~meV) when comparing with 
denser meshes. For the calculation of electron-phonon (e-ph) coupling matrix 
elements, a denser $k$-point mesh of $48 \times 48 \times 48$ for the Brillouin 
sampling was required to achieve convergence. The reported numerical parameters 
were chosen such that frequencies were converged better than $0.05$~meV, 
constituting a compromise between precision and computational cost.
\\

The standard DFT spin-polarized (ssp) calculation tends to overestimate the 
ordered magnetic moment in the FM state especially for itinerant magnets, where 
spin fluctuations effectively weaken the magnetic interaction. An elegant 
method to amend this is the spin scaling exchange-correlation (ssxc) correction 
\cite{ortz,onou,sharma}. This scheme introduces a linear scaling of the 
magnetization density and potential in the evaluation of the 
exchange-correlation potential by a single scaling parameter ($s$), while the 
charge potential remains unchanged. 
It can be interpreted as a scaling of the effective Stoner parameter, which 
characterizes the strength of the magnetic interaction \cite{ortz}. This scheme 
is variational and adapts the spin polarization for all standard types of 
exchange-correlation functionals in a self-consistent way.
A value $s=1$ corresponds to the standard ssp calculation. This approach has 
previously been applied to the Mn$_{1-x}$Fe$_x$Si solid solution \cite{khan}, 
where it successfully described a significant reduction of the ordered magnetic 
moment, while the $\mathrm{LDA}+U$ method required unrealistic high values of 
$U > 6-7$~eV to achieve the same reduction \cite{colly,dutta}.

\section{Results and Discussion}

\subsection{Structural properties}

\begin{figure}
\begin{center}
\includegraphics[trim = 4mm 45mm 40mm 7mm, clip, width=0.50 \textwidth]{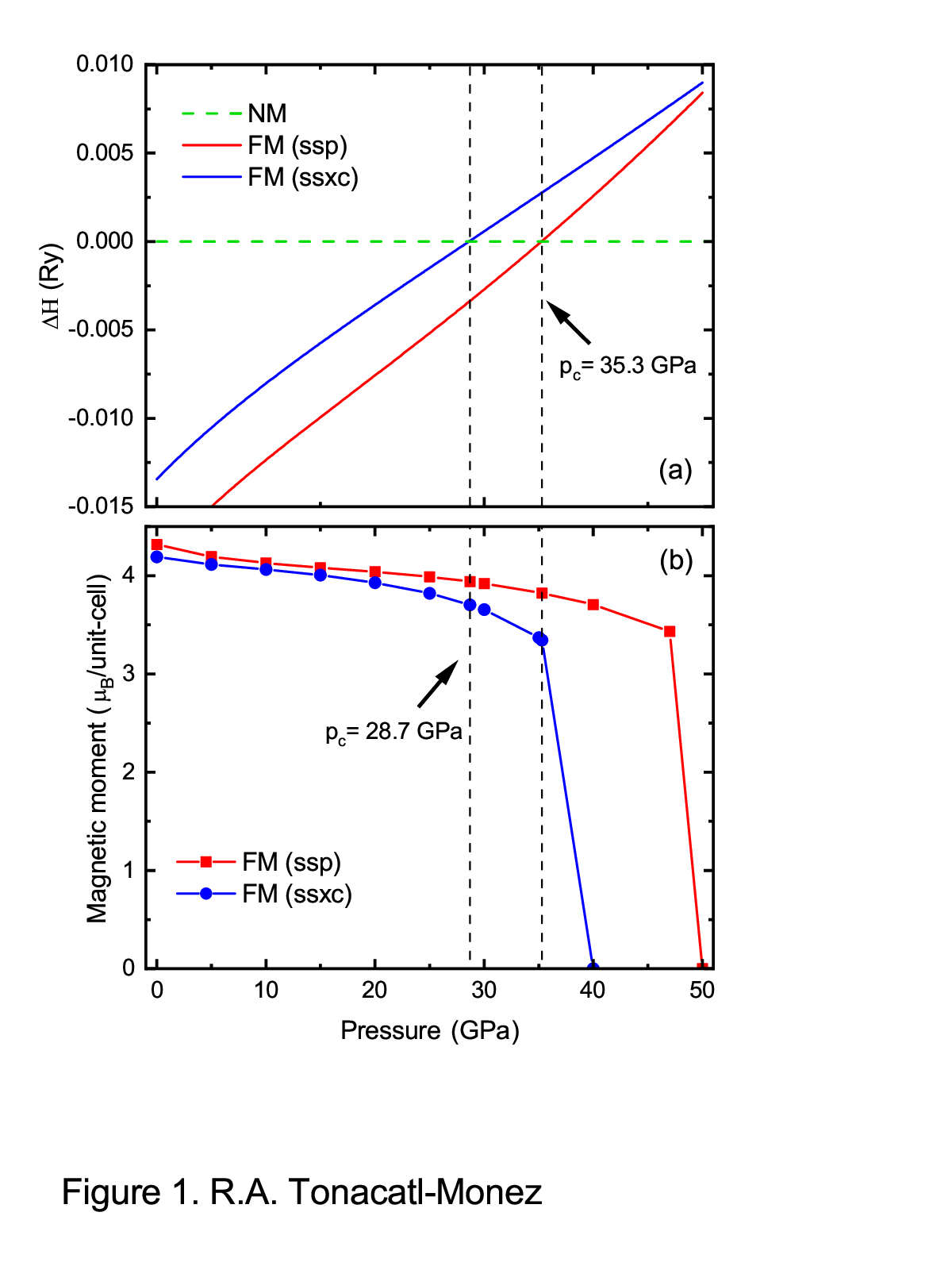}
\caption{Evolution of (a) the enthalpy difference (between NM and FM phases) as 
a function of pressure, where a negative value indicates a more stable magnetic 
phase, and (b) magnetic moment for the FM ssp and ssxc schemes. The vertical 
lines indicate the critical pressure ($p_c$): $28.7$~GPa for ssxc with 
$s=0.979$, and $35.3$~GPa for ssp.} \label{FIG01}
\end{center}
\end{figure}

The B20 crystal structure is cubic with four formula units (f.u.) within the 
unit cell, where its Wyckoff atomic positions ($4a$) are given by the 
coordinates $(u, u, u)$, $(0.5+u, 0.5-u, -u)$, $(-u, 0.5+u, 0.5-u)$ and 
$(0.5-u, -u, 0.5+u)$ \cite{lebe}. The FeGe lattice structure was fully 
optimized by determining the total energy for a set of volumes. For a given 
volume, the internal coordinates $u_{\mathrm{Fe}}$ and $u_{\mathrm{Ge}}$ were 
optimized by force minimization, with a force threshold of $0.005$~Ry/a.u. 
Results for $E(V)$ were then fitted to the Birch-Murnaghan equation of state 
(EOS) \cite{BM}, from which the optimized structure as well as the relation 
$p(V)$ were extracted. Optimized structural parameters for the NM and FM phases 
are shown in Table \ref{TAB01}, along with the corresponding magnetic moment 
($m$) per unit cell. Values for the FM state do agree nicely with available 
experimental data \cite{pshen,pedrz,lebe}, but the magnetic moment is 
overestimated by about 10\%.

\begin{table*}[ht]
\caption{The optimized structural parameters, in particular the volume ($V_0$), 
bulk modulus ($B_0$), and lattice internal parameters ($u$) for Fe and Ge, for 
the NM and FM phases at ambient pressure. Shown is also the magnetic moment $m$ 
(per unit cell) without applied pressure and the critical pressure $p_c$ for 
the standard spin-polarized case (ssp) and for the spin-scaling 
exchange-correlation correction (ssxc) with $s=0.979$. The last column 
indicates the calculated magnetic moment at $p_c$.}
\label{TAB01}
\begin{center}
\begin{tabular}{c | c | c | c |c | c | c | c}
\br 
  & $V_0$      & $B_0$  & $u$   & $u$  & $m$       & $p_c$ & $m @ p_c$ \\
  & (Bohr$^3$) & (GPa)  & (Fe)  & (Ge) & ($\mu_B$) & (GPa) & ($\mu_B$) \\
\mr 
NM      & 677.48 &	170.23 & 0.13335 & 0.83842 & $-$  &	$-$  & $-$  \\
FM-ssp	& 687.99 &  149.43 & 0.13495 & 0.84178 & 4.31 & 35.3 & 3.82 \\
FM-ssxc & 686.85 &	156.74 & 0.13493 & 0.84173 & 4.19 &	28.7 & 3.70 \\
\mr
Exp.	& 700.63 \cite{pshen} & 147.00 \cite{pshen} & 0.13524 \cite{lebe} & 0.8414 \cite{lebe} 
& 3.928 \cite{spen} & 28.5 \cite{barla} & $-$ \\
\br 
\end{tabular}
\end{center}
\end{table*}

The critical pressure ($p_c$) where the magnetic transition from FM to NM takes 
place, is determined by comparing the enthalpy $H(p)=E(V)+pV$ for both phases 
as a function of pressure. In Figure \ref{FIG01}(a) the enthalpy difference 
($\Delta H$) between the FM and NM phases is presented. For low pressures, 
$\Delta H$ is negative indicating the stability of the FM state. The sign 
change of $\Delta H$ signals the phase transition from FM to NM as function of 
pressure and determines $p_c$. From this analysis, $p_c=35.3$~GPa was 
determined for the ssp scheme, which differs considerably from the experimental 
value of $28.5$~GPa \cite{barla}. \\

\begin{figure} 
\begin{center}
\includegraphics[trim = 5mm 130mm 0mm 5mm, clip, width=0.50 \textwidth]{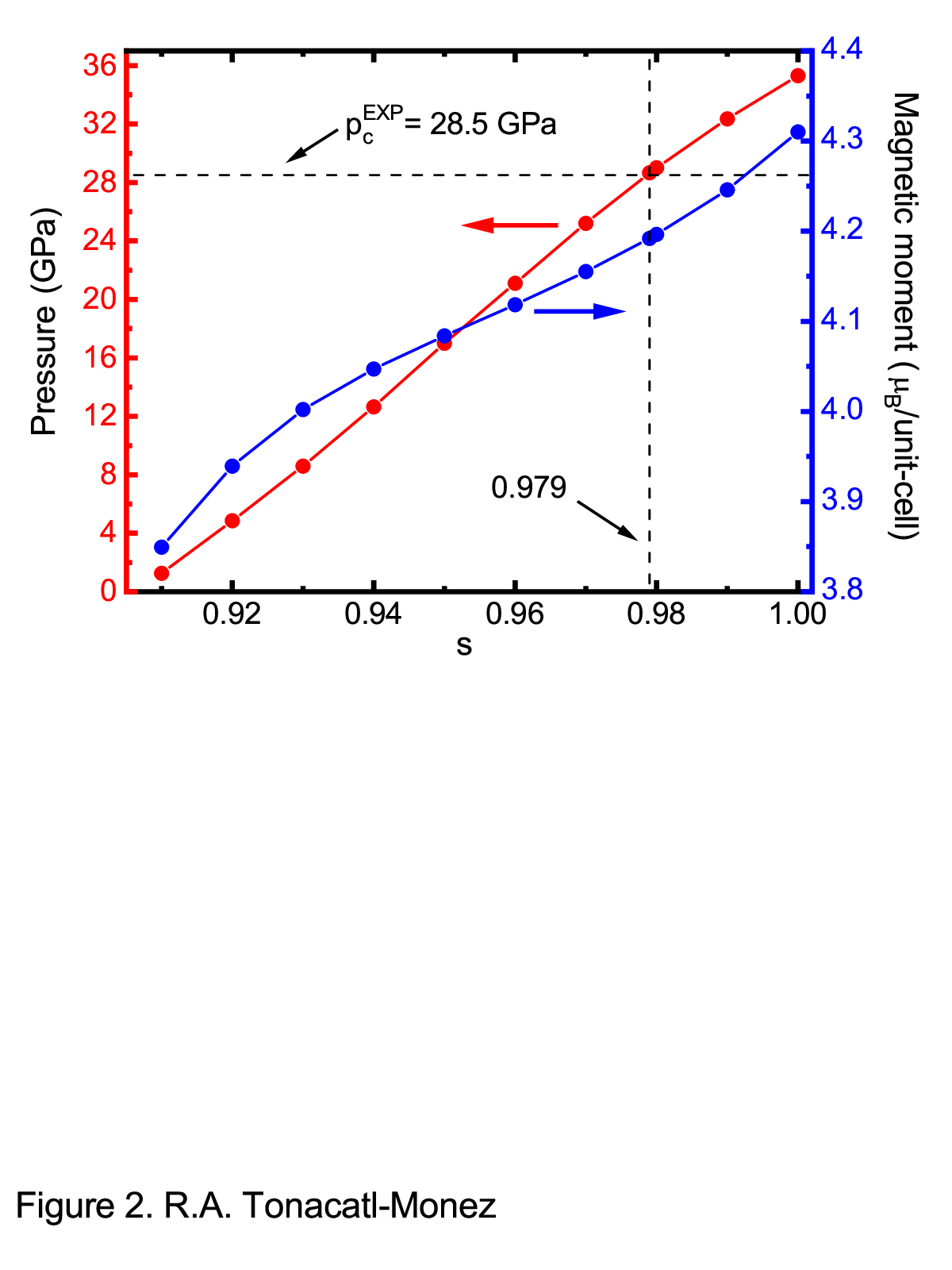}
\caption{Determined critical pressure ($p_c$) and magnetic moment as a function 
of the ssxc parameter $s$. The dotted horizontal line indicates the experimental 
$p_c$ value \cite{barla}.} 
\label{FIG02}
\end{center}
\end{figure}

In order to improve the critical-pressure value, the spin scaling 
exchange-correlation (ssxc) correction was employed \cite{ortz,onou,sharma}. 
Figure \ref{FIG02} shows the dependence of $p_c$ and the magnetic moment $m$ at 
$p=0$~GPa on the scaling parameter $s$. The critical pressure is very sensitive 
to $s$, reaching the experimental value with a moderate reduction to $s=0.979$. 
Simultaneously, the magnetic moment for $p=0$~GPa is also improved, showing a 
difference of roughly 6$\%$ with respect to the experimental value (see 
Table\ref{TAB01}). \\

The evolution of the magnetic moment as a function of pressure obtained by both 
schemes, ssp and ssxc (with $s=0.979$), is compared in Figure \ref{FIG01}(b). 
Trends are qualitatively similar, but with a reduced pressure range for the FM 
phase and with slightly smaller magnetic moments in the case of the ssxc sheme. 
It is important to mention that there are still magnetic solutions for FeGe at 
pressure values above $p_c$ for both schemes. However, as can be seen from the 
enthalpy difference in Figure \ref{FIG01}(a), these FM solutions are 
energetically less favorable than the NM state, representing metastable states. 
Thus the pressure induced transition from FM to NM is predicted to be of first 
order. Interestingly, the structural properties obtained by the ssxc 
scheme with $s=0.979$ differ only marginally from those obtained by the ssp 
scheme, as can be seen in Table \ref{TAB01}, indicating that both schemes 
equally well describe the experimental data.

\subsection{Electronic properties}

\begin{figure} 
\begin{center}
\includegraphics[trim = 4mm 32mm 0mm 5mm, clip, width=0.50 \textwidth]{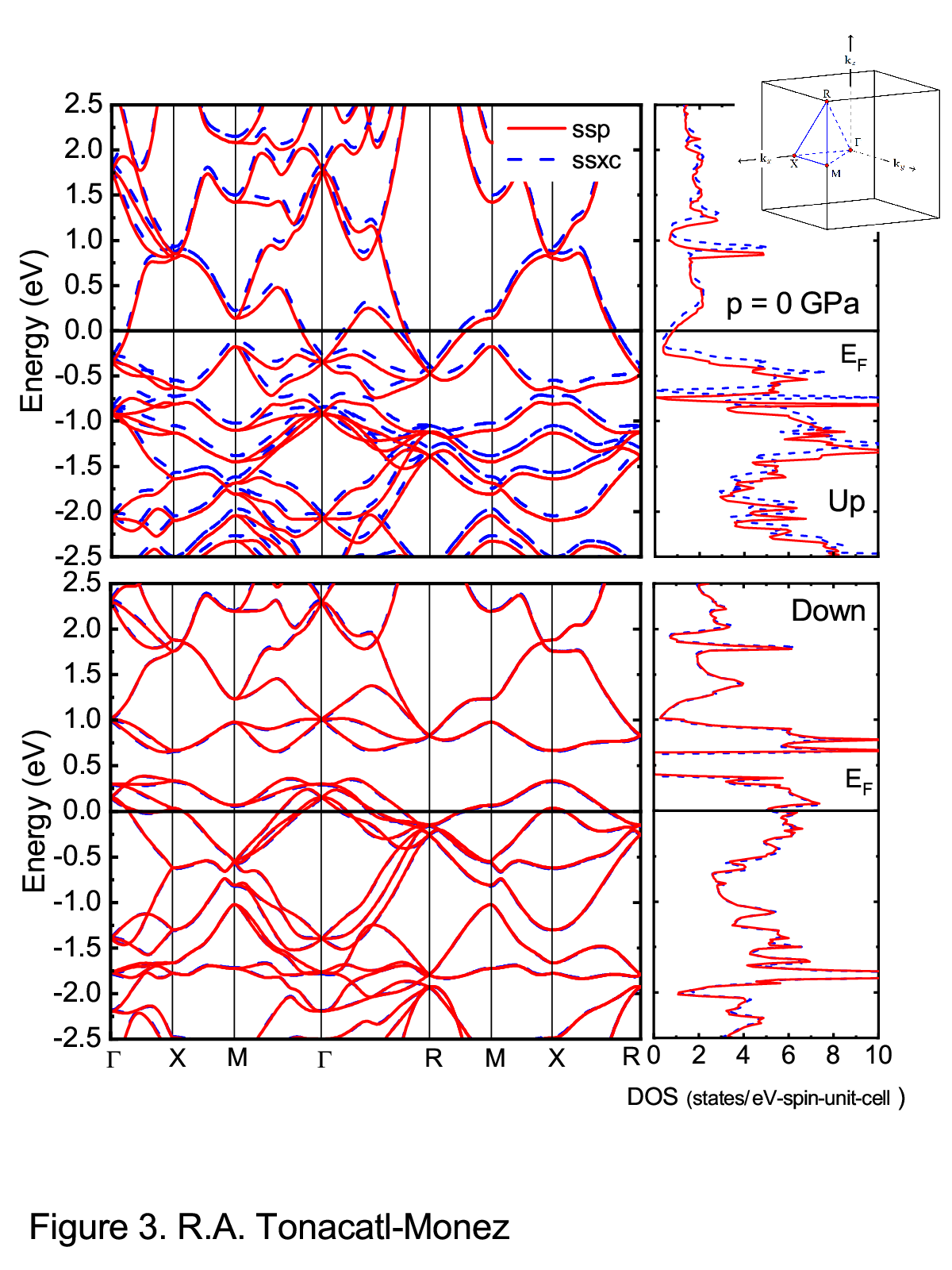}
\caption{Electronic band structure and density of states of FM FeGe 
($p=0$~GPa), for both spin-channels, comparing ssp and ssxc (with $s=0.979$) 
schemes. The related first Brillouin of the simple cubic structure zone is 
presented as an inset.} \label{FIG03}
\end{center}
\end{figure}

\begin{figure}
\begin{center}
\includegraphics[trim = 5mm 55mm 0mm 5mm, clip, width=0.50 \textwidth]{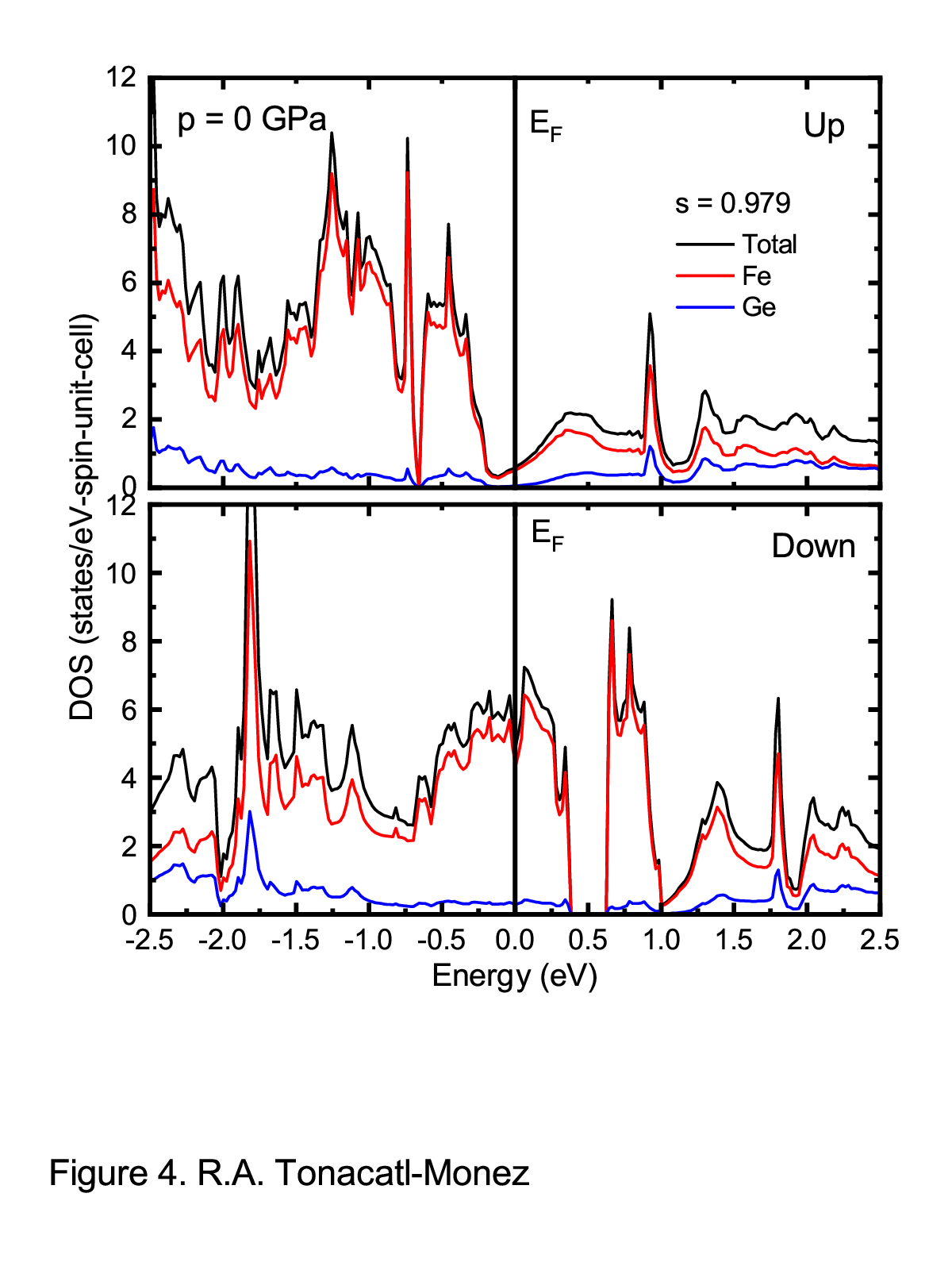}
\caption{Spin-polarized density of states (DOS) for FM FeGe, obtained with the 
ssxc scheme (using $s=0.979$) without applied pressure, showing the 
contribution by atom (in the unit cell).} \label{FIG04}
\end{center}
\end{figure}

The band structure and density of states (DOS) for FM FeGe, without applied 
pressure ($p=0$~GPa), are shown in Figure \ref{FIG03} for the spp and ssxc with 
$s=0.979$, respectively, to elucidate the effects of spin-scaling on its 
electronic properties. In general, the spin-down channel possesses a higher 
value of the density of states at the Fermi level, $N(E_F)$, as compared to the 
spin-up channel. In particular, the spin-down channel does not show any visible 
effects when using the ssxc approach, while that is not the case for the 
spin-up channel, where ssxc leads to a slight shift in the bands and DOS 
towards higher energies than ssp. 
Since the ssxc scheme scales down the exchange potential and reduces the 
magnetic moment, it tends to reduce the original exchange splitting between 
spin channels. Due to the larger $N(E_F)$ for the spin-down channel as compared 
to the spin-up channel, the spin-down channel essentially pins the Fermi 
energy, while the spin-up bands shift relative to $E_F$. 
Atom-resolved DOS for the FM state of FeGe at ambient pressure are presented in 
Figure \ref{FIG04} for the ssxc scheme, using $s=0.979$. For both spin 
channels, the larger contribution is due to Fe-$d$ orbitals, with much lower 
participation of Ge states, mainly coming from $d$ and $p$ orbitals.

\begin{figure}
\begin{center}
\includegraphics[trim = 5mm 50mm 0mm 5mm, clip, width=0.50 \textwidth]{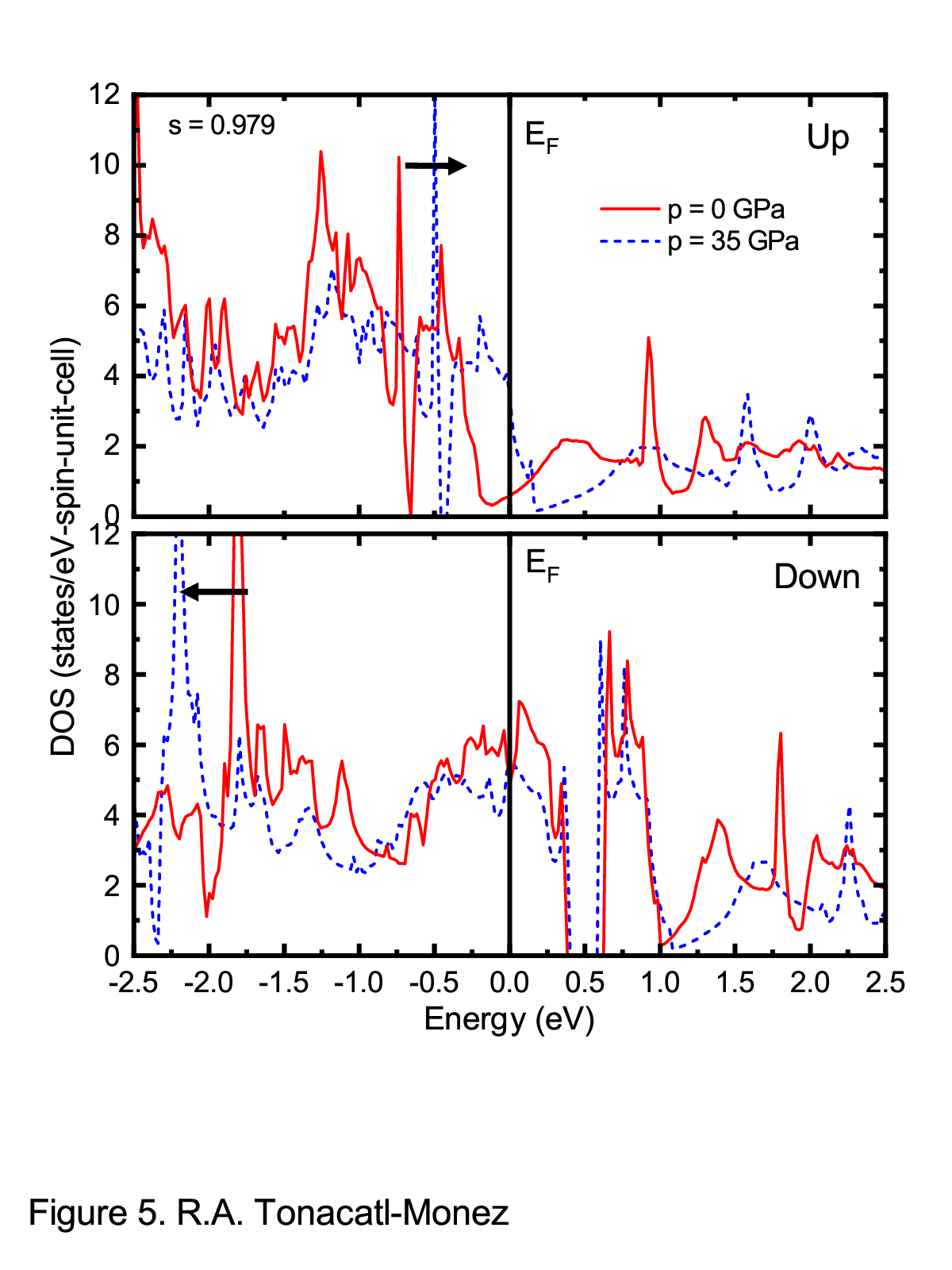}
\caption{Spin-polarized DOS of two different applied pressure values, $0$~GPa 
and $35$~GPa, calculated under the ssxc scheme with $s=0.979$.} 
\label{FIG05}
\end{center}
\end{figure}

To illustrate the effect of pressure on the electronic structure, Figure 
\ref{FIG05} presents the DOS for two different pressures, calculated with ssxc 
($s=0.979$). For the spin-up channel, there is a shift toward higher energy 
values, while the opposite is observed for the spin-down channel, that is, a 
shift to lower energy values with respect to the Fermi level, especially for 
states below $1$~eV. This behavior is understood by the tendency to reduce the 
magnetic moment of FM FeGe and to approach the non-magnetic (NM) solution as 
the pressure increases.

\begin{figure}
\begin{center}
\includegraphics[trim = 5mm 55mm 0mm 5mm, clip, width=0.50 \textwidth]{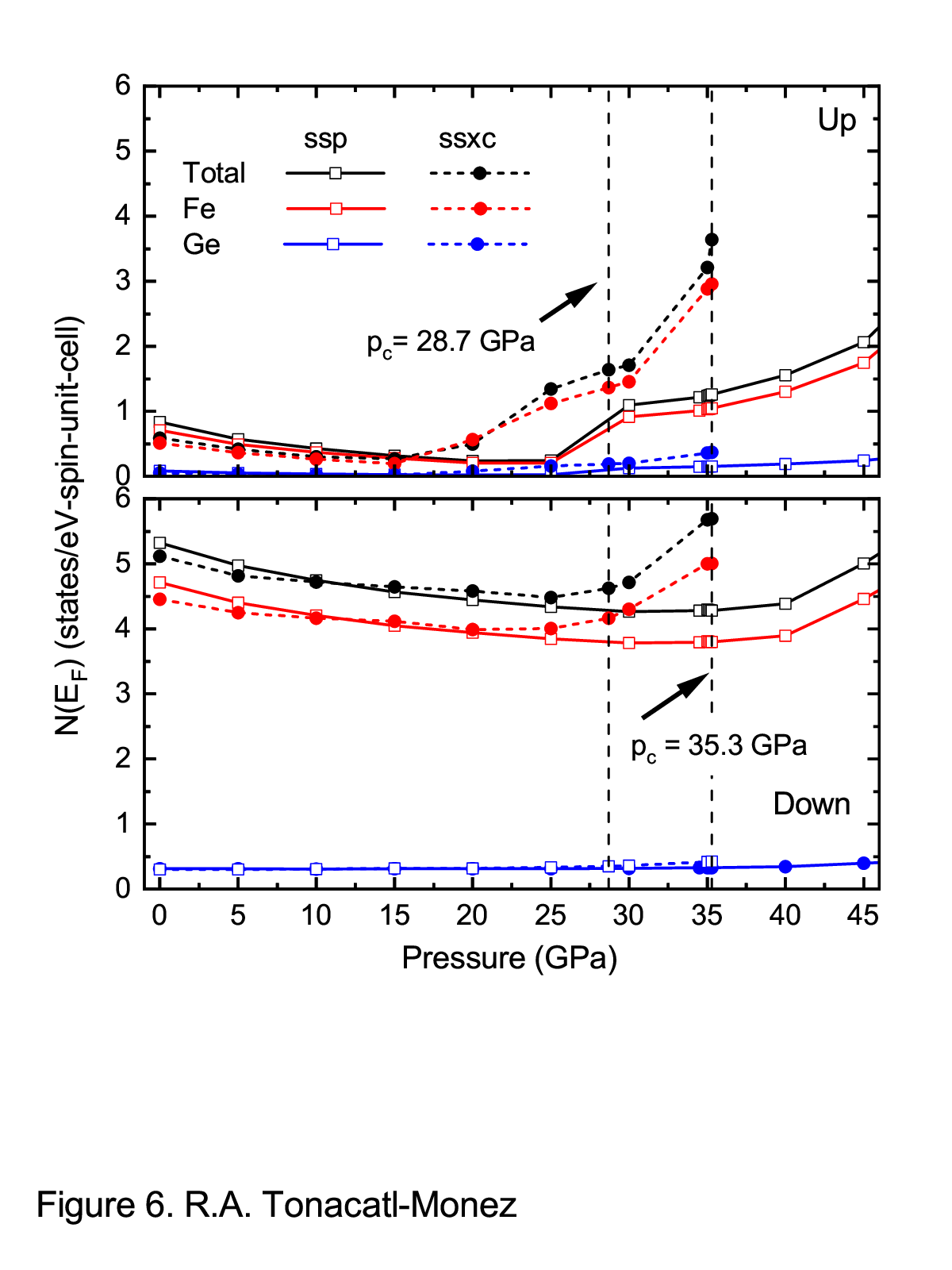}
\caption{Evolution of the density of states at the Fermi level, $N(E_F)$, as a 
function of applied pressure for FM FeGe, separated by atomic contribution and 
calculated with both schemes, ssp and ssxc ($s=0.979$). The dotted vertical 
lines represent the critical pressure value obtained for each scheme: 
$28.7$~GPa for ssxc, and $35.3$~GPa for ssp.} 
\label{FIG06}
\end{center}
\end{figure}

The evolution of the density of states at the Fermi level ($N(E_F)$) as a 
function of pressure, for each spin channel, is presented in Figure \ref{FIG06} 
for both schemes, ssp and ssxc ($s=0.979$). In general, both schemes show not 
only the same trend as a function of pressure, but also agree quantitatively 
for a wide range of applied pressure. Differences start to appear as one 
approaches $p_c$. The smaller $p_c$ value for ssxc results in a shrinking of 
the pressure range. However, independent of the spin channel, the $N(E_F)$ 
differences between the schemes are not very large, even for pressures close to 
$p_c$, indicating that spin-scaling induces only subtle changes in the 
electronic properties.

\subsection{Lattice dynamical properties and electron-phonon (e-ph) coupling}

\begin{figure} 
\begin{center}
\includegraphics[trim = 0mm 175mm 0mm 10mm, clip, width=0.70 \textwidth]{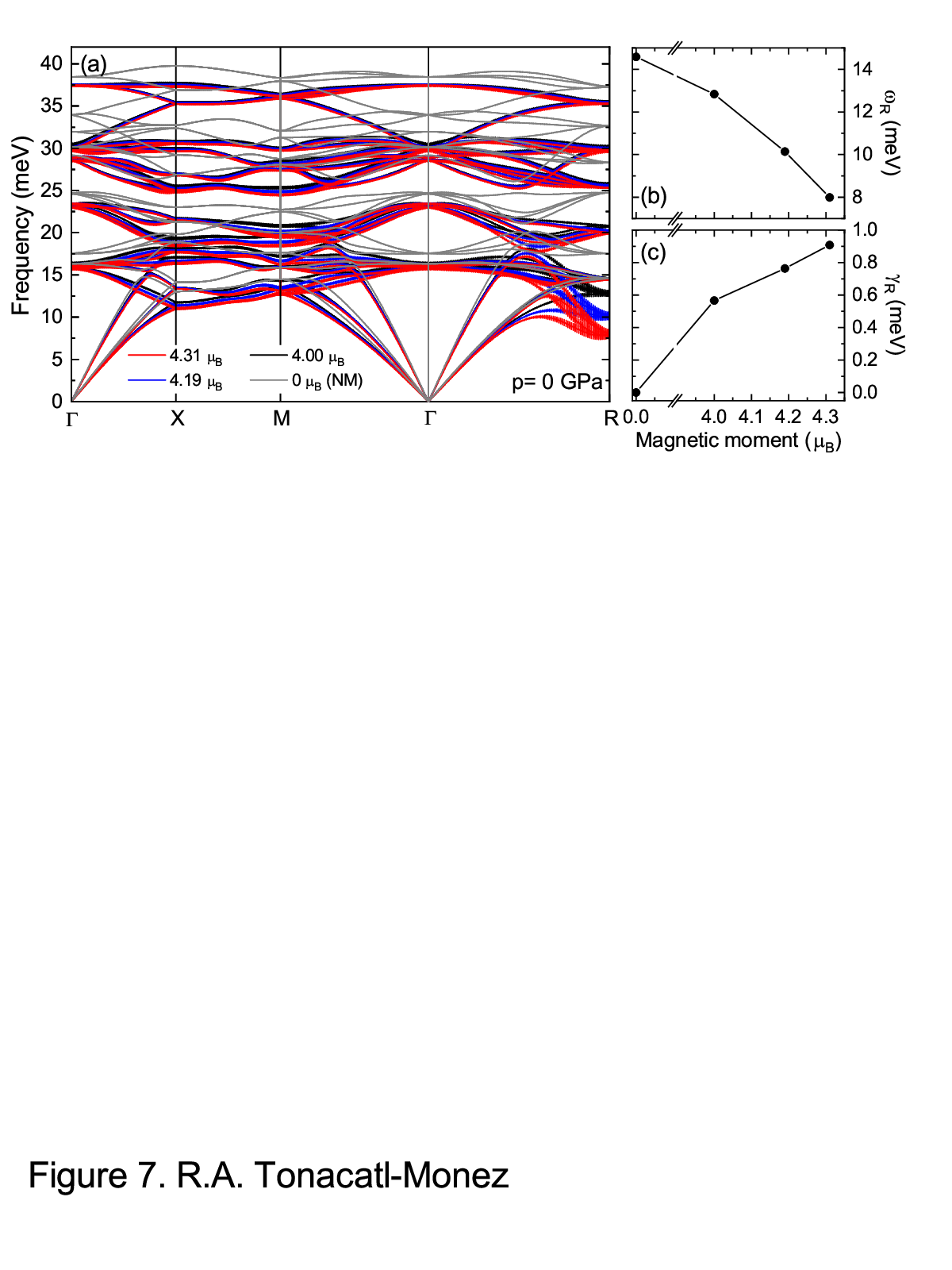}
\caption{(a) Comparison of phonon dispersion and linewidths (phonon branches 
width), without applied pressure, for several magnetic cases: $4.31~\mu_B$ 
(ssp), $4.19~\mu_B$ ($s=0.979$), $4.0~\mu_B$ ($s=0.930$), and 
$0~\mu_B$ (NM). Evolution as a function of magnetic moment of (b) frequency and 
(c) linewidth for the lowest-frequency mode at the $R$-point.} \label{FIG07}
\end{center}
\end{figure}

To study the sensitivity of phonons and e-ph coupling with respect to the 
magnetic order, in particular with respect to the magnetic moment, phonon 
dispersion and linewidths were calculated for four cases with different 
magnetic moments, obtained by changing the $s$ parameter: $4.31~\mu_B$ 
($s=1$), $4.19~\mu_B$ ($s=0.979$), $4.0~\mu_B$ ($s=0.930$), and the NM case 
($0~\mu_B$) for comparison. 
The first one corresponds to the standard spin-polarized calculation (ssp), and 
the second one to the $s$ value that fits the critical pressure to the 
experimental data. The third one uses an even smaller $s$ value to further 
suppress the magnetic moment, while the last one corresponds to the non-spin 
polarized case. All calculations were done for their corresponding
optimized structures at $p=0$~GPa.
The ssp scheme shows a phonon anomaly close to the $R$-point in the acoustic 
branches, as previously noted \cite{wilhm1}, while the corresponding linewidths 
are, by far, the largest ones in the whole phonon spectra. With decreasing 
magnetic moment, the frequencies of all phonons increase, but the most 
prominent modification is a drastic hardening of the low-frequency acoustic 
phonon branches in the neighborhood of the $R$-point, accompanied by a 
reduction of their linewidths. In particular, the lowest-frequency mode at the 
$R$-point exhibits a clear correlation between phonon hardening and linewidth 
reduction (see Figs. \ref{FIG07}(a) and \ref{FIG07}(b)).
Branches attached to this mode also show the highest sensitivity to changes in 
the ordered magnetic moment. For example, when going from the $s=1$ to the 
$s=0.979$ case, the phonon anomaly as well as the corresponding linewidths are 
visibly downsized, despite an only $3\%$ difference between their corresponding 
magnetic moments. \\

\begin{figure}
\begin{center}
\includegraphics[trim = 0mm 52mm 5mm 7mm, clip, width=1.05 \textwidth]{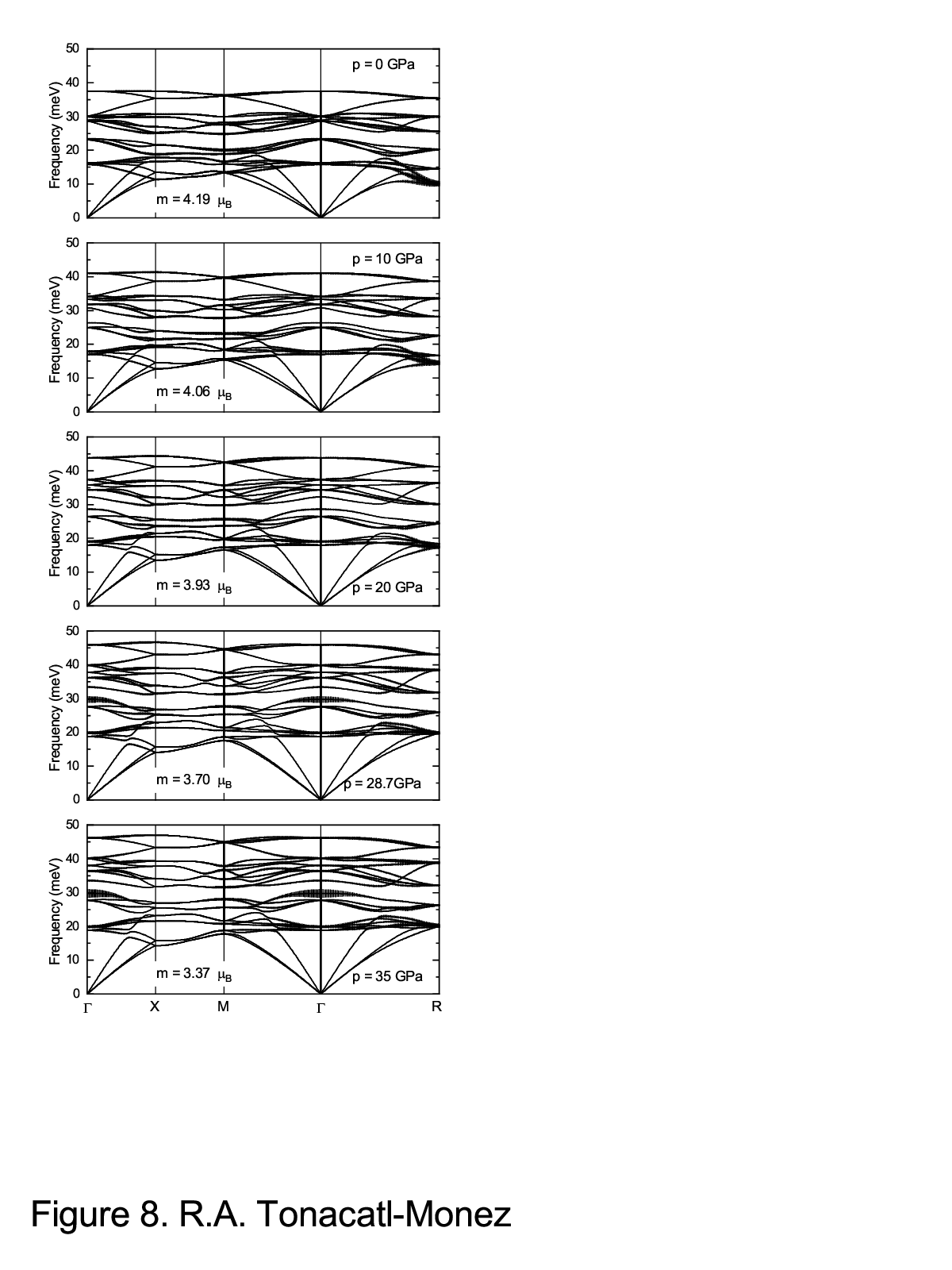}
\caption{Phonon dispersion and linewidths calculated for the FM ssxc scheme 
($s=0.979$) at several applied pressure values.} \label{FIG08}
\end{center}
\end{figure}

Phonon dispersions and linewidths for the FM phase with a scaling parameter 
$s=0.979$ are presented for several  pressure values in Figure \ref{FIG08}. 
Increasing pressure generally hardens the phonon frequencies, which is 
particularly evident for the acoustic branches in the $\Gamma-R$ path where the 
phonon anomaly is present at $p=0$~GPa. This phonon renormalization is notably 
stronger in the pressure range where the FM phase is the more stable one, that 
is, for pressure values lower than $p_c=28.7$~GPa. \\

\begin{figure}
\begin{center}
\includegraphics[trim = 0mm 160mm 0mm 7mm, clip, width=0.60 \textwidth]{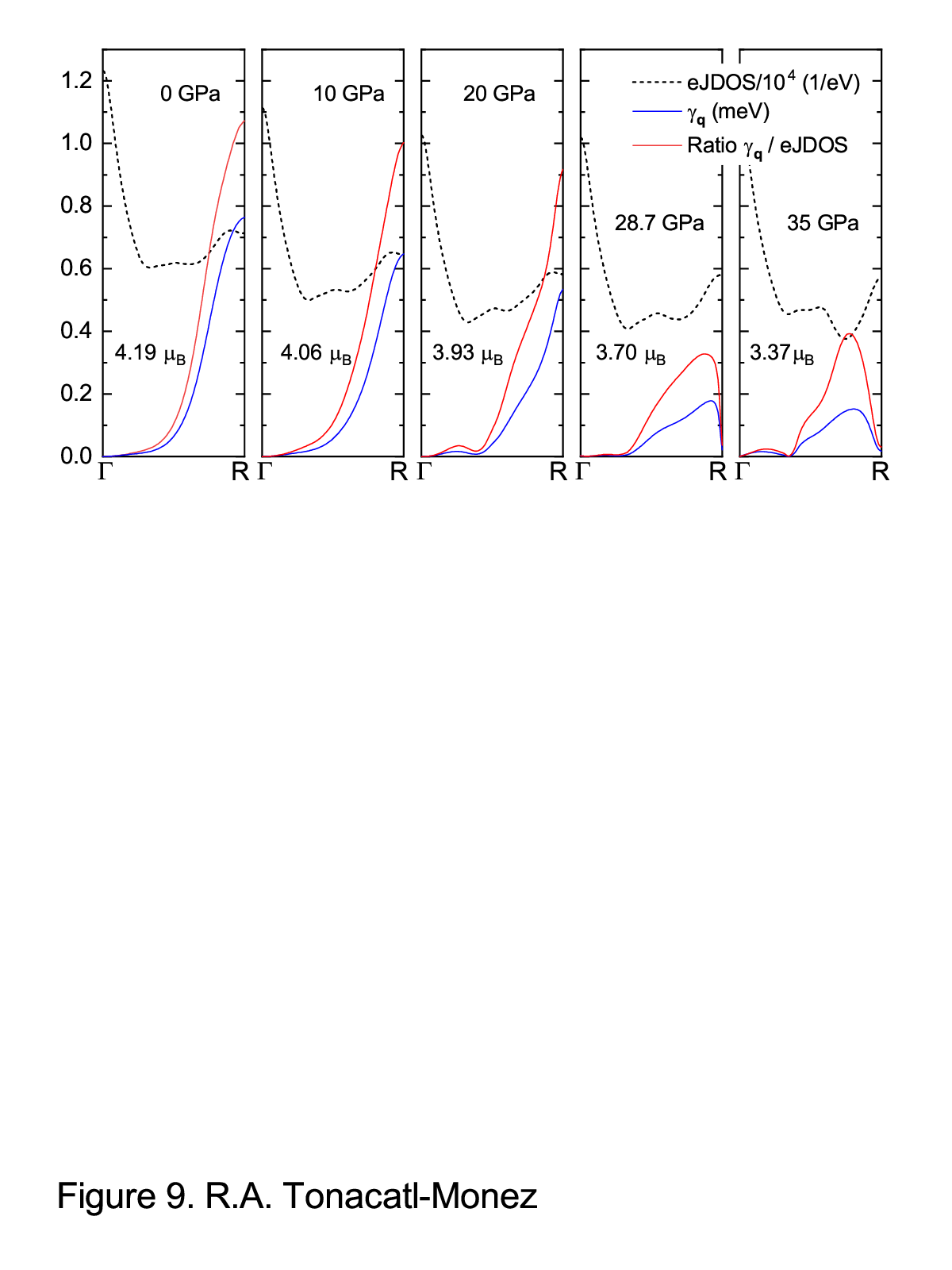}
\caption{Comparison of eJDOS (scaled), phonon linewidth of the 
lowest-frequency acoustic branch $\gamma_\mathbf{q}$, and the ratio of 
$\gamma_{\mathbf{q}}$ and eJDOS, which describes an average of the 
$\mathbf{q}$-dependent e-ph coupling matrix elements, for several pressure 
values, calculated under the ssxc scheme with $s=0.979$.} \label{FIG09}
\end{center}
\end{figure}

For the analysis of the phonon linewidth, $\gamma_{\mathbf{q}\eta}$, we recall 
that it is expressed as

\begin{equation}
\gamma_{\mathbf{q}\eta} = 2\pi\omega_{\mathbf{q}\eta}\sum_{\mathbf{k}\nu\nu'}
\left|g^{\mathbf{q}\eta}_{\mathbf{k}+\mathbf{q}\nu',\mathbf{k}\nu}\right|^2
\delta\left(\epsilon_{\mathbf{k}\nu}-E_F\right)\delta\left(\epsilon_{\mathbf{k}+\mathbf{q}\nu'}-E_F\right),
\end{equation}

\noindent
where $\omega_{\mathbf{q}\eta}$ is the frequency of the phonon mode at the 
$\mathbf{q}$ vector and branch $\eta$, and $\epsilon_{\mathbf{k}\nu}$ is the 
one-electron band energy with momentum $\mathbf{k}$ and band index $\nu$.  
$\gamma_{\mathbf{q}\eta}$ is closely related to the electronic joint density of 
states (eJDOS), given by $\sum_{\mathbf{k}\nu\nu'}\delta\left(\epsilon_{\mathbf{k}\nu}-E_F\right)\delta\left(\epsilon_{\mathbf{k}+\mathbf{q}\nu'}-E_F\right)$, 
but with the difference that the sum is weighted by squared electron-phonon 
coupling matrix elements 
$g^{\mathbf{q}\eta}_{\mathbf{k}+\mathbf{q}\nu',\mathbf{k}\nu}$. %
We explicitly calculated eJDOS (which is also referred as the Fermi surface 
nesting function) to discriminate between the influence of electron-phonon coupling
matrix elements and Fermi surface geometry on the linewidths. The same $k$-mesh 
and broadening parameters were used as in the linewidth calculation. Since the 
FeGe ground state is spin polarized, the eJDOS is also spin dependent. However,
due to its low $N(E_F)$, {the contribution from the} spin-up channel is almost 
negligible in comparison with the spin-down channel (a couple of orders of 
magnitude smaller, not shown here). Thus, we present only the results for the 
spin-down channel in Figure \ref{FIG09} together with the linewidth of the 
corresponding lowest-frequency acoustic branch in the $\Gamma$-$R$ path. 
Additionally, the ratio between $\gamma_{\mathbf{q}}$ and eJDOS is presented, 
which defines an average of the momentum-dependent e-ph coupling matrix 
elements. \\

From the $p=0$~GPa case (see Figure \ref{FIG09}), it is clear that the 
linewidth is significantly larger as $\mathbf{q}$ approaches the $R$-point. As 
pressure increases, with a simultaneous decrease of the magnetic moment, the 
linewidth starts to decrease and practically disappears for $p \geq p_c$. 
Although eJDOS shows a slight shoulder around the $R$-point throughout the 
analyzed pressure range, it is unclear whether the linewidth behavior is 
related to nesting features. However, the $\gamma_\mathbf{q}/\mathrm{eJDOS}$ 
ratio, which, as stated before, is a measure of the $\mathbf{q}$-dependent 
e-ph coupling matrix elements, closely follows the same behavior as the 
linewidth: a steady increase as $\mathbf{q}$ approaches the $R$-point, and a 
drastic reduction as the applied pressure increases and the magnetic moment 
decreases.
Thus, large linewidth values around the $R$-point for the lowest-frequency 
acoustic branch are linked to presence of magnetism, and originates in a 
significant increase of e-ph coupling matrix elements in this region.

\section{Conclusion}

As a summary, we have performed a first-principles study on the structural, 
electronic, lattice dynamical properties, and electron-phonon coupling of 
ferromagnetic cubic FeGe as a function of applied pressure. Implementing the 
spin-scaling exchange-correlation (ssxc) approach, allowed us to adjust the 
critical pressure, $p_c$, to the experimental value. A small reduction of the 
scaling parameter from $s=1$ to $s=0.979$ was sufficient to reduce the magnetic 
moment and the energetics of the ferromagnetic phase to yield a proper $p_c$ 
value. The spin-scaling correction essentially affects the energetic position 
of the spin-up bands only. 
Subsequently, using the ssxc scheme, at a fixed scaling parameter $s=0.979$, we 
analyzed the evolution of the phonon dispersion, $\omega_{\mathbf{q}}$, and 
linewidths, $\gamma_\mathbf{q}$, as a function of applied pressure. Firstly, we 
observed that the pronounced softening and large linewidths at the $R$-point, 
obtained by standard DFT calculations, are lessened by the ssxc scheme, which 
reduces at the same time the magnetic moment. Secondly, the phonon anomaly and 
associated large linewidth values diminished with increasing pressure, 
practically disappearing beyond $p_c$. 
To understand the origin of this behavior, we compared the eJDOS, 
$\gamma_\mathbf{q}$, and the average of the momentum-dependent e-ph coupling matrix 
elements (obtained through the $\gamma_\mathbf{q}/\mathrm{eJDOS}$ ratio) for 
the lowest-frequency acoustic branch in the $\Gamma$-$R$ path. With increasing 
pressure, we found very similar trends between $\gamma_\mathbf{q}$ and this 
ratio as $\mathbf{q}$ approaches the $R$-point, with a simultaneous reduction 
of the magnetic moment. 
These findings strongly suggest that the observed correlation between the 
decrease in the magnetic moment and the reduction in linewidths with applied 
pressure has its origin in significant changes of the electron-phonon coupling 
matrix elements. This represents a mechanism distinct to other members of the 
B20 family, such as the Mn$_{1-x}$Fe$_x$Si solid solution, where 
doping induced changes in linewidths are dominated by changes in the 
Fermi surface geometry, i.e. nesting. 

\section{Data availability statement}

The data that support the findings of this research are available upon 
reasonable request from the authors.

\section{Acknowledgments}

This research was partially supported by Vicerrector\'ia de Investigaci\'on y 
Estudios de Posgrado (VIEP), Benem\'erita Universidad Aut\'onoma de Puebla 
(BUAP) under Grant No. 100517450-VIEP2025, the Deutscher Akademischer Austausch 
Dienst (DAAD), and the Karlsruher Institut f\"ur Technologie (KIT), Germany. 
One of the authors (R.A.T.M.) gratefully acknowledges financial support from 
the Secretar\'ia de Ciencia, Humanidades, Tecnolog\'ia e Innovaci\'on (Secihti, 
M\'exico).

\section*{References}

\bibliography{fege_pressure_biblio}

\providecommand{\newblock}{}
\begin{thebibliography}{10}
\expandafter\ifx\csname url\endcsname\relax
  \def\url#1{{\tt #1}}\fi
\expandafter\ifx\csname urlprefix\endcsname\relax\def\urlprefix{URL }\fi
\providecommand{\eprint}[2][]{\url{#2}}

\bibitem{kise}
Kiselev N, Bogdanov A, Sch{\"a}fer R and R{\"o}{\ss}ler U 2011 {\em Phys. D
  Appl. Phys.\/} {\bf 44} 392001

\bibitem{kana1}
Kanazawa N 2015 {\em Charge and Heat Transport Phenomena in Electronic and Spin
  Structures in B20-type Compounds\/} (Springer Japan)

\bibitem{eo}
Eo Y~S, Avers K, Horn J~A, Yoon H, Saha S~R, Suarez A, Fuhrer M~S and Paglione
  J 2023 {\em Applied Physics Letters\/} {\bf 122} 233102

\bibitem{ishi}
Ishikawa Y, Tajima K, Bloch D and Roth M 1976 {\em Solid State
  Communications\/} {\bf 19} 525--528

\bibitem{grigo1}
Grigoriev S~V, Siegfried S~A, Altynbayev E~V, Potapova N~M, Dyadkin V, Moskvin
  E~V, Menzel D, Heinemann A, Axenov S~N, Fomicheva L~N and Tsvyashchenko A~V
  2014 {\em Phys. Rev. B\/} {\bf 90}(17) 174414

\bibitem{wang}
Whang K, Wei W and Du H 2025 {\em Adv. Funct. Mater.\/} {\bf 35} 2416203

\bibitem{kana2}
Kanazawa N, Shibata K and Tokura Y 2016 {\em New Journal of Physics\/} {\bf 18}
  045006

\bibitem{schle}
Schlesinger Z, Fisk Z, Zhang H~T, Maple M, DiTusa J and Aeppli G 1993 {\em
  Phys. Rev. Lett.\/} {\bf 71} 1748

\bibitem{buschi}
Buschinger B, Geibel C, Steglich F, Mandrus D, Young D, Sarrao J and Fisk Z
  1997 {\em Phys. B\/} {\bf 230} 784

\bibitem{jacca}
Jaccarino V, Wertheim G, Wernick J, Walker L and Arajs S 1993 {\em Phys.
  Rev.\/} {\bf 160} 476

\bibitem{mandru}
Mandrus D, Sarrao J, Migliori A, Thompson J and Fisk Z 1995 {\em Phys. Rev.
  B\/} {\bf 51} 4763

\bibitem{menzel}
Menzel D, Popovich P, Kovaleva N~N, Schoenes J, Doll K and Boris A~V 2009 {\em
  Phys. Rev. B\/} {\bf 79}(16) 165111

\bibitem{delaire}
Delaire O, Marty K, Stone M, Kent P, Lucas M, Abernathy D, Mandrus D and Sales
  B 2011 {\em Proceedings of the National Academy of Sciences\/} {\bf 108}
  4725--4730

\bibitem{krann}
Krannich S, Sidis Y, Lamago D, Heid R, Mignot J~M, L$\rm{\ddot{o}}$hneysen H,
  Ivanov A, Steffens P, Keller T, Wang L, Goering E and Weber F 2015 {\em Nat
  Commun\/} {\bf 6} 8961

\bibitem{bannen}
Bannenberg L, Dalgliesh R, Wolf T, Weber F and Pappas C 2018 {\em Phys. Rev.
  B\/} {\bf 98} 184431

\bibitem{khan}
Khan N, De~la Pe\~na Seaman O, Heid R, Voneshen D, Said A, Bauer A, Konrad T,
  Merz M, Wolf T, Pfleiderer C and Weber F 2024 {\em Phys. Rev. B\/} {\bf 109}
  184306

\bibitem{wilhm1}
Wilhelm H, Leonov A, R$\rm{\ddot{o}}${\ss}ler U, Burger P, Hardy F, Meingast C,
  Gruner M, Schnelle W, Schmidt M and Baenitz M 2016 {\em Phys. Rev. B\/} {\bf
  94} 144424

\bibitem{stolt}
Stolt M, Sigelko X, Mathur N and Jin S 2018 {\em Chem. Mater.\/} {\bf 30} 1146

\bibitem{yu}
Yu X, Kanazawa N, Onose Y, Kimoto K, Zhang W, Ishiwata S, Matsui Y and Tokura Y
  2011 {\em Nat. Mater.\/} {\bf 10} 106

\bibitem{spen}
Spencer C, Gayles J, Porter N, Sugimoto S, Aslam Z, Kinane C, Charlton T,
  Freimuth F, Chadon S, Langridge S, Sinova J, Felser C, Bl$\rm{\ddot{u}}$gel
  S, Mokrousov Y and Marrows C 2018 {\em Phys. Rev. B\/} {\bf 97} 214406

\bibitem{wilhm2}
Wilhelm H, Baenitz M, Schmidt M, R$\rm{\ddot{a}}${\ss}ler U, Leonov A and
  Bogdanov A 2011 {\em Phys. Rev. Lett.\/} {\bf 107} 127203

\bibitem{barla}
Barla A, Wilhelm H, Forthaus M, Strohm C, R{\"u}ffer R, Schmidt M, Koepernik K,
  R{\"o}{\ss}ler U and Abd-Elmeguid M 2015 {\em Phys. Rev. Lett.\/} {\bf 114}
  016803

\bibitem{neef}
Neef M, Doll K and Zwicknagl G 2009 {\em Phys. Rev. B\/} {\bf 80} 035122

\bibitem{hohen}
Hohenberg P and Kohn W 1964 {\em Phys. Rev.\/} {\bf 136} B864

\bibitem{kohn}
Kohn W and Sham L~J 1965 {\em Phys. Rev.\/} {\bf 140} A1133

\bibitem{mbpp}
Meyer B, Els$\rm{\ddot{a}}$sser C, Lechermann F and F$\rm{\ddot{a}}$hnle M
  \normalfont{FORTRAN90 Program for Mixed-Basis Pseudopotential Calculations
  for Crystals, Max-Planck-Institut f$\rm{\ddot{u}}$r Metallforschung,
  Sttutgart (unplublished)}

\bibitem{perdw}
Perdew J, Burke K and Ernzerhof M 1996 {\em Phys. Rev. Lett.\/} {\bf 77} 3865

\bibitem{lebe}
Lebech B, Bernhard J and Freltoft T 1989 {\em J . Phys.: Condens. Matter\/}
  {\bf 1} 6105

\bibitem{vand}
Vanderbilt D 1985 {\em Phys. Rev. B\/} {\bf 32} 8412

\bibitem{louie}
Louie S, Ho K and Cohen M 1979 {\em Phys. Rev. B\/} {\bf 19} 1774

\bibitem{baron}
Baroni S, Gironcoli S~D and Corso A~D 2001 {\em Phys. Rev. Lett.\/} {\bf 73}
  515

\bibitem{heid2}
Heid R and Bohnen K~P 1999 {\em Phys. Rev. B\/} {\bf 60} R3709

\bibitem{ortz}
Ortenzi L, Mazin I~I, Blaha P and Boeri L 2012 {\em Phys. Rev. B\/} {\bf 86}
  064437

\bibitem{onou}
Onourah I, Bonf\'a P and Renzi R~D 2018 {\em Phys. Rev. B\/} {\bf 97} 174414

\bibitem{sharma}
Sharma S, Gross E, Sanna A and Dewhurst J 2018 {\em Chem. Theory Comput.\/}
  {\bf 14} 1247

\bibitem{colly}
Collyer R and Browne D~A 2008 {\em Phys. B\/} {\bf 403} 1420

\bibitem{dutta}
Dutta P and Pandey S 2018 {\em Comput. Condens. Matter\/} {\bf 16} e00325

\bibitem{BM}
Birch F 1947 {\em Phys. Rev.\/} {\bf 71} 809

\bibitem{pshen}
Pshenay-Severin D and Burkov A 2019 {\em Materials\/} {\bf 17} 12

\bibitem{pedrz}
Pedrazzini P, Wilhelm H, Jaccard D, Jarlborg T, Schmidt M, Hanfland M, Akselrud
  L, Yuan H~Q, Schwarz U, Grin Y and Steglich F 2007 {\em Phys Rev Lett.\/}
  {\bf 98} 047204

\end{thebibliography}


\end{document}